\documentstyle[thmsa,a4,11pt,sw20lart]{article}

\input{tcilatex}
\begin{document}

\author{C. Bagnuls\thanks{%
Service de Physique de l'Etat Condens\'{e}} \ and C.\ Bervillier\thanks{%
Service de Physique Th\'{e}orique e-mail: bervil@spht.cea.saclay.fr} \\
CE Saclay, F91191 Gif-sur-Yvette Cedex, France}
\title{Renormalization group domains of the scalar Hamiltonian.\thanks{%
Based on a talk given at ``RG\ 2000'', Taxco, Mexico, January 1999.}}
\date{\today}
\maketitle

\begin{abstract}
Using the local potential approximation of the exact renormalization group
(RG) equation, we show the various domains of values of the parameters of
the $O(1)$-symmetric scalar Hamiltonian. In three dimensions, in addition to
the usual critical surface $S_{\text{c}}$ (attraction domain of the
Wilson-Fisher fixed point), we explicitly show the existence of a
first-order phase transition domain $S_{\text{f}}$ separated from $S_{\text{c%
}}$ by the tricritical surface $S_{\text{t}}$ (attraction domain of the
Gaussian fixed point). $S_{\text{f}}$ and $S_{\text{c}}$ are two distinct
domains of repulsion for the Gaussian fixed point, but $S_{\text{f}}$ is not
the basin of attraction of a fixed point. $S_{\text{f}}$ is characterized by
an endless renormalized trajectory lying entirely in the domain of negative
values of the $\varphi ^{4}$-coupling. This renormalized trajectory exists
also in four dimensions making the Gaussian fixed point ultra-violet stable
(and the $\varphi _{4}^{4}$ renormalized field theory asymptotically free
but with a wrong sign of the perfect action). We also show that very
retarded classical-to-Ising crossover may exist in three dimensions (in fact
below four dimensions). This could be an explanation of the unexpected
classical critical behavior observed in some ionic systems.

\end{abstract}

\section{Introduction\label{Intro}}

The object of this paper is to carry on studying the local potential
approximation of the exact renormalization group (RG) equation for the
scalar theory \cite{4595}. In a previous publication \cite{3554} (to be
considered as the part I of the present work), we had already considered this
approximation with a view to qualitatively discuss the connection between
the standard perturbative renormalization of field theory (as it can be
found in most textbooks on field theory, see for example \cite{3533}) and
the modern view \cite{440} in which the renormalized parameters of a field
theory are introduced as the ``relevant'' directions of a fixed point (FP)
of a RG transform. Actually the local potential approximation, which allows
us to consider all the powers of the field $\varphi $ on the same footing,
is an excellent textbook example of the way infinitely many degrees of
freedom are accounted for in (nonperturbative) RG theory. Almost all the
characteristics of the RG theory are involved in this approximation. The
only lacking features are related to phenomena highly correlated to the non
local parts neglected in the approximation and when the critical exponent $%
\eta $ is small (especially for $d=4$ and $d=3$), one expects the
approximation to be qualitatively correct on all aspects of the RG theory 
\cite{4595}.

In the following we look at the domains of attraction or of repulsion of
fixed points in the $O(1)$ scalar theory in three and four dimensions ($d=3$
and $d=4$).

At first sight, one could think that the issue considered is very simple
since, with regard to criticality, the $O(1)$-symmetric systems in three
dimensions are known to belong to the same class of universality (the Ising
class). Now, because the Ising class is associated to the domain of
attraction of the unique (non-trivial) Wilson-Fisher fixed point \cite{439},
then by adjusting one parameter (in order to reach the critical temperature%
\footnote{%
We assume that the second relevant field, corresponding to the magnetic
field for magnetic systems, is set equal to zero.}) any $O(1)$ scalar
Hamiltonian should be driven to the Wilson-Fisher fixed point under the
action of renormalization. Consequently there would have only two domains
for the $O(1)$ scalar theory: the critical subspace $S_{\text{c}}$ (of
codimension 1) in the Wilson space ($S$) of infinite dimensions of the
Hamiltonian parameters (in which the RG transforms generate flows) and the
complement to $S$ of $S_{\text{c}}$ (corresponding to noncritical
Hamiltonians).

In fact, that is not correct because there is another fixed point in $S$:
the Gaussian fixed point which, although trivial, controls tricritical
behaviors in three dimensions. Now each FP has its own basin of attraction
in $S$ \cite{439}. The attraction domain of the Gaussian FP is the
tricritical subspace $S_{\text{t}}$ of codimension 2 (with no intersection
with $S_{\text{c}}$). In addition, we show that there is a second subspace
of codimension 1 in $S$, called $S_{\text{f}}$, which is different
from $S_{\text{c}}$, and thus which is not a domain of attraction to the
Wilson-Fisher fixed point. There is no FP to which a point of $S_{\text{f}}$
is attracted to. $S_{\text{f}}$ is characterized by a negative sign of the $%
\varphi ^{4}$-Hamiltonian parameter $u_{4}$ and is associated with systems
undergoing a first-order phase transition \cite{4612}. We show that an
endless attractive RG trajectory is associated to this domain of first-order
transitions. It is a renormalized trajectory (denoted below by T$_{1}^{"}$)
that emanates from the Gaussian fixed point. The frontier between $S_{\text{f%
}}$ and $S_{\text{c}}$ corresponds to the tricritical subspace $S_{\text{t}}$
which is the domain of attraction of the Gaussian fixed point while $S_{%
\text{f}}$ and $S_{\text{c}}$ are two distinct domains of repulsion for the
Gaussian fixed point.

Actually, the situation is conform to the usual view. Considering the famous 
$\varphi ^{4}$-model [Landau-Ginzburg-Wilson (LGW) Hamiltonian] in which the
associated coupling $u_{4}$ is positive, then the Hamiltonian at criticality
is attracted exclusively to the Wilson-Fisher fixed point, but if $u_{4}$ is
negative, a $\varphi ^{6}$-term is required for stability, but then one may
get either a tricritical phase transition or a second- or a first-order
transition \cite{3390}. In the present study we do not truncate the
Hamiltonian which involves all the powers of the field $\varphi $.

We explicitly show that a system which would correspond to an initial point
lying very close to the frontier $S_{\text{t}}$ in the critical side (in $S_{%
\text{c}}$) would display a retarded classical-to-Ising crossover \cite{4003}%
. This result is interesting with regard to ionic systems (for example) in
which a classical behavior has been observed while an Ising like critical
behavior was expected. The eventuality of a retarded crossover from the
classical to the Ising behavior has previously been mentioned but without
theoretical explanation on how this kind of crossover could develop \cite
{3575}. In \cite{4003} a calculation suggests that the RPM model for ionic
systems would specifically correspond to a scalar Hamiltonian with a
negative sign for the $\varphi ^{4}$-Hamiltonian parameter (but the order
parameter chosen is not the bulk density \cite{4244}). That calculation has
motivated the present study.

We also indicate that the renormalized trajectory T$_{1}^{^{\prime \prime }}$
still exists in four dimensions. This makes the Gaussian fixed point
ultraviolet stable and the scalar field theory {\em formally\/}
asymptotically free. However the associated ``perfect'' action \cite{4283}
would have the wrong sign to provide us with an acceptable (well defined)
field theory.

The organisation of the paper is as follows. In section \ref{LPA-ERGE} we
briefly present the local potential approximation of the exact RG equation
to be studied. We introduce the strategy we have chosen to solve the
resulting nonlinear differential equation with a view to show the
trajectories of interest in the space $S$ of infinite dimension. Because the
practical approach to the Gaussian fixed point is made difficult due to the
logarithmic slowness characteristic of a marginally irrelevant direction
(for $d=3$), we found it useful to first test our numerical method with a
close approach to the Wilson-Fisher fixed point. We present the
characteristic results of this approach and the various kinds of domains
corresponding to $u_{4}>0$ (somewhat a summary of \cite{3554})

In section \ref{Uneg} we describe the various kinds of attraction or
repulsion domains of the Gaussian fixed point (for a negative value of the $%
\varphi ^{4}$-Hamiltonian parameter) corresponding to tricritical, critical
and first-order subspaces. Then we discuss the consequences and especially
explicitly show how a retarded crossover from the classical to the Ising
behavior can be obtained.

We then shortly discuss the case $d=4$ when $u_{4}<0$.

In two appendices we report on some technical aspects of the numerical
treatment of the RG equation studied, in particular on the appearance of
spurious nontrivial tricritical fixed points (Appendix \ref{Spurious}).

\section{The RG equation studied\label{LPA-ERGE}}

The local potential approximation has been first considered by Nicoll et al 
\cite{3480} from the sharp cutoff version of the exact RG equation of Wegner
and Houghton \cite{414}, it has been rederived by Tokar \cite{4398} by using
approximate functional integrations and rediscovered by Hasenfratz and
Hasenfratz \cite{2085}. As in \cite{3554} we adopt the notation of the
latter authors and consider the following nonlinear differential equation
for the simple function $f(l,\varphi )$: 
\begin{equation}
\dot{f}=\frac{K_{d}}{2}\frac{f^{\prime \prime }}{1+f^{\prime }}+\left( 1-%
\frac{d}{2}\right) \varphi f^{\prime }+\left( 1+\frac{d}{2}\right) f
\label{eq4}
\end{equation}
in which a prime refers to a derivative with respect to the constant
dimensionless field $\varphi $ (at constant $l$) and $f(l,\varphi
)=V^{\prime }(l,\varphi )$ is the derivative of the dimensionless potential $%
V(l,\varphi )$; $\dot{f}$ stands for $\partial f/\partial l|_{\varphi }$ in
which $l$ is the scale parameter defined by $\frac{\Lambda }{\Lambda _{0}}=$e%
$^{-l}$ and corresponding to the reduction to $\Lambda $ of an arbitrary
initial momentum scale of reference $\Lambda _{0}$ (the initial sharp
momentum cutoff). Finally, $K_{d}$ is the surface of the $d$-dimensional
unit sphere divided by $\left( 2\pi \right) ^{d}$.

A fixed point is solution of the equation $\dot{f}=0$. The study of the
resulting second order differential equation provides the following results:

\begin{itemize}
\item  $d\geq 4$, no FP is found except the Gaussian fixed point.

\item  $3\leq d<4$, one nontrivial FP (the Wilson-Fisher fixed point \cite
{439}) is found \cite{2085,38,3816}

\item  A new nontrivial FP emanates from the origin (the Gaussian fixed
point) below each dimensional threshold $d_{k}=2k/(k-1)$, $k=2,3,\ldots
,\infty $ \cite{2625}.
\end{itemize}

If one represents the function $f(l,\varphi )$ as a sum of monomials of the
form:

\[
f(l,\varphi )=\sum_{n}u_{2n}\left( l\right) \varphi ^{2n-1} 
\]
then, for $d=3$, the Wilson-Fisher fixed point $f^{*}$ is located in $S$ at 
\cite{3554}: $u_{2}^{*}=-0.461533\cdots $, $u_{4}^{*}=3.27039\cdots $, $%
u_{6}^{*}=14.4005\cdots $, $u_{8}^{*}=32.31289\cdots $, etc.

Once the FP is known, one may study its vicinity which is characterized by
orthogonal directions corresponding to the infinite set of eigenvectors,
solutions of the differential equation (\ref{eq4}) linearized at $f^{*}$.
The eigenvectors associated to positive eigenvalues are said relevant; when
the eigenvalues are negative they are said irrelevant and marginal otherwise 
\cite{413}.

The relevant eigenvectors correspond to directions along which the RG
trajectories go away from the FP and the irrelevant eigenvectors correspond
to directions along which the trajectories go into the FP. A marginal
eigenvector may be relevant or irrelevant.

Our present FP $f^{*}$ has only one relevant direction and infinitely many
irrelevant directions (no marginal direction, however see \cite{4595}). As
already explained and shown in \cite{3554}, in order to approach $f^{*}$
starting from an initial point in $S$, one must adjust one parameter of the
initial function $f(0,\varphi )$. This amounts to fixing the temperature of
a system to its critical temperature.

Starting with a known initial function (at ``time'' $l=0$) say:

\[
f(0,\varphi )=u_{2}\left( 0\right) \varphi +u_{4}\left( 0\right) \varphi
^{3} 
\]
we adjust $u_{2}\left( 0\right) $ to the critical value $%
u_{2}^{c}=-0.29958691\cdots $ corresponding to $u_{4}\left( 0\right) =3$ so
that $f(l,\varphi )$ [solution at time $l$ of the differential equation (\ref
{eq4})] approaches $f^{*}$ when $l\rightarrow \infty $. The approach to $%
f^{*}$ is characterized by the least negative eigenvalue $\lambda
_{2}=-1/\omega _{1}$ ($\omega _{1}$ was noted $\omega $ in \cite{3554}).
This means that, in the vicinity of $f^{*}$ any parameter $u_{n}\left(
l\right) $ evolves as follows ($l\rightarrow \infty $): 
\[
u_{n}\left( l\right) \simeq u_{n}^{*}+a_{n}\exp \left( -\omega _{1}l\right) 
\]

Fig. \ref{fig1} illustrates this feature for the first four $u_{n}\left(
l\right) $'s in the approach to $f^{*}$. In \cite{3554} the two associated
attractive trajectories (locally tangent to the least irrelevant eigenvector
in the vicinity of $f^{*}$) was noted T$_{1}$ and T$_{1}^{\prime }$.

One may also constrain the trajectory to approach $f^{*}$ along the second
irrelevant direction (with the associated attractive trajectories noted T$%
_{2}$ or T$_{2}^{\prime }$ in \cite{3554} and associated with the second
least negative eigenvalue $\lambda _{3}=-1/\omega _{2}$). In this case a
second parameter of the initial $f$ must be adjusted, e.g., $u_{4}\left(
0\right) $ must be adjusted to $u_{4}^{c}$ {\em and} {\em simultaneously} $%
u_{2}\left( 0\right) $ to the corresponding $u_{2}^{c}$, see \cite{38,3554}.
Then, in the vicinity of $f^{*}$, any parameter $u_{n}\left( l\right) $ will
evolve as follows: 
\[
u_{n}\left( l\right) \cong u_{n}^{*}+a_{n}^{\prime }\exp \left( -\omega
_{2}l\right) 
\]

Looking for this kind of approach to $f^{*}$, we have found that $%
6.66151663<u_{4}^{c}<6.66151669$ and $u_{2}^{c}=-0.58328898880579\cdots $
This has allowed us to estimate $\omega _{2}\cong 2.84$. Although the
shooting method is certainly not well adapted to the determination of the
eigenvalues (see the huge number of digits required in the determination of $%
u_{2}^{c}$ and $u_{4}^{c}$), our estimate is close to $\omega _{2}\cong
2.8384$ found by Comellas and Travesset \cite{3860}.

Because $u_{4}^{c}$ cannot be perfectly determined, the trajectory leaves
the trajectory T$_{2}$ before reaching $f^{*}$ to take one of the two
directions T$_{1}$ or T$_{1}^{^{\prime }}$ (corresponding to $\omega _{1}$).
Fig \ref{fig2} illustrates this effect with the evolution, for $n=2$, of the
following effective eigenvalue:

\begin{equation}
\omega _{\text{eff}}^{(n)}\left( l\right) =-\frac{d^{2}u_{n}(l)/dl^{2}}{%
du_{n}(l)/dl}  \label{eq-OmegaEff}
\end{equation}
the definition of which does not refer explicitly to $f^{*}$. The evolution
of $\omega _{\text{eff}}\left( l\right) $ shows a flat extremum (or a flat
inflexion point) at an RG eigenvalue of $f^{*}$ each time the RG flow runs
along an eigendirection in the vicinity of $f^{*}$.

Similarly to $u_{4}^{c}$, the value $u_{2}^{c}$ cannot be perfectly
determined, consequently the trajectory ends up going away from the fixed
point. This provides us with the opportunity of determining the only
positive (the relevant) eigenvalue corresponding to the critical exponent $%
\nu =\lambda _{1}=-1/\omega _{0}$ [$\omega _{\text{eff}}\left( l\right) $
shows then a flat extremum at $\omega _{0}$ when the flow still runs in the
close vicinity of $f^{*}$]. Finally, far away from the fixed point, the RG
trajectory approaches the trivial high temperature fixed point characterized
by a classical eigenvalue (equal to $\frac{1}{2}$). The global picture
summarizing the evolution of $\omega _{\text{eff}}\left( l\right) $ along a
RG trajectory initialized in such a way as to approach $f^{*}$ first along T$%
_{2}$, is drawn on fig. \ref{fig3}.

The values we have determined by this shooting method are (for eigenvalues
other than $\omega _{2}$ already mentioned):

\begin{eqnarray*}
\omega _{1} &\cong &0.5953 \\
\nu &\cong &0.68966
\end{eqnarray*}
which are close to the values found, for example, in \cite{2085,3860}: $%
\omega _{1}\cong 0.5952$ and $\nu \cong 0.6895$.

\section{Trajectories for $u_{4}<0$ \label{Uneg}}

In the preceding section, we have obtained a RG trajectory approaching the
Wilson-Fisher fixed point $f^{*}$ along T$_{2}$ by adjusting two parameters
of the initial Hamiltonian ($u_{4}^{c}$ and $u_{2}^{c}$). This is exactly
the procedure one must follow to determine a tricritical RG trajectory
approaching the Gaussian fixed point in three dimensions (because of its two
relevant directions). The only difficulty is to discover initial points in $%
S $ which are attracted to the Gaussian fixed point. To this end, we again
use the shooting method.

From usual arguments on the LGW Hamiltonian and from the work done by
Aharony on compressible ferromagnets \cite{4612}, one expects to find the
tricritical surface in the sector $u_{4}<0$ (and with $u_{2}>0$). Thus we
have tried to approach the Gaussian fixed point starting with initial
function $f(0)$ of the form: 
\begin{equation}
f(0,\varphi )=u_{2}\left( 0\right) \varphi +u_{4}\left( 0\right) \varphi
^{3}+u_{6}(0)\varphi ^{5}  \label{eq-Init-Tri1}
\end{equation}
with (not large) negative values of $u_{4}\left( 0\right) $, for example $%
u_{4}\left( 0\right) =-1$.

Because the Gaussian fixed point is twice unstable, we must adjust two
parameters to approach it starting with (\ref{eq-Init-Tri1}). We do that by
successive tries (shooting method). For example, if we choose $u_{4}\left(
0\right) =-1$ and $u_{6}(0)=3$ and determine a value of $u_{2}\left(
0\right) $ such as to get a trajectory which does not go immediately
towards the trivial high temperature fixed point, the best we obtain is a
trajectory which approaches the Wilson-Fisher fixed point (thus the
corresponding initial point belongs to the attraction domain of $f^{*}$
although $u_{4}\left( 0\right) <0$ \cite{4612}). But if $u_{6}(0)=2$, the
adjustment of $u_{2}\left( 0\right) $ with a view to counterbalance the
effect of the most relevant direction of the Gaussian fixed point (which
would drive the trajectory toward the high temperature FP) yields a runaway
RG flow towards larger and larger negative values of $u_{4}\left( l\right) $%
. From now on, the target is bracketted: the tricritical trajectory
corresponding to $u_{4}\left( 0\right) =-1$ can be obtained with a value of $%
u_{6}(0)$ in the range $\left] 2,3\right[ $ (we actually find a rather close
approach to the Gaussian fixed point for $2.462280>u_{6}(0)>2.4622788$ and $%
6.4618440\cdots >u_{2}(0)>6.4618407\cdots $).

In order to understand the origin of the direction of runaway in the sector
of negative values of $u_{4}$, it is worth to study the properties of the
Gaussian fixed point by linearization of the RG flow equation in the
vicinity of the origin. If we request the effective potential to be bounded
by polynomials then the linearization of eq. (\ref{eq4}) identifies with the
differential equation of Hermite's polynomials of degree $n=2k-1$ for the
set of discrete values of $\lambda $ satisfying \cite{2085}:

\begin{equation}
\frac{2+d-2\lambda _{k}}{d-2}=2k-1\text{ \qquad }k=1,2,3,\cdots  \label{vp}
\end{equation}
from which it follows that

\begin{itemize}
\item  for $d=4$: $\lambda _{k}=4-2k$ \qquad $k=1,2,3,\cdots $, there are
two non-negative eigenvalues: $\lambda _{1}=2$ and $\lambda _{2}=0$

\item  for $d=3$: $\lambda _{k}=3-k$ \qquad $k=1,2,3,\cdots $, there are
three non-negative eigenvalues: $\lambda _{1}=2$, $\lambda _{2}=1$ et $%
\lambda _{3}=0$
\end{itemize}

If we denote by $\chi _{k}(\varphi )$ the eigenfunctions associated to the
eigenvalue $\lambda _{k}$, it comes:

\begin{itemize}
\item  $\chi _{1}^{+}=\varphi $, $\chi _{2}^{+}=\varphi ^{3}-\frac{3}{2}%
\varphi $, $\chi _{3}^{+}=\varphi ^{5}-5\varphi ^{3}+\frac{15}{4}\varphi $, $%
\cdots $, whatever the spatial dimensionality $d$.
\end{itemize}

The upperscript ``$+$'' is just a reminder of the fact that the
eigenfunctions are defined up to a global factor and thus the functions $%
\chi _{k}^{-}(\varphi )=-\chi _{k}^{+}(\varphi )$ are also eigenfunctions
with the same eigenvalue $\lambda _{k}$.

\subsection{Case $d=3$}

Similarly to $\chi _{2}^{+}$, the direction provided by $\chi _{2}^{-}$ in $%
S $ is a direction of instability of the Gaussian fixed point. Now $\chi
_{2}^{+}$ is associated with the well known renormalized trajectory T$_{1}$
on which is defined the usual (massless) $\varphi _{3}^{4}$-field theory 
\cite{38,3554}, for the same reasons a renormalized trajectory T$%
_{1}^{^{\prime \prime }}$ locally tangent to $\chi _{2}^{-}$ in the vicinity
of the origin of $S$ exists with the same properties as T$_{1}$ (see \cite
{3554}). The difference is that T$_{1}^{^{\prime \prime }}$ lies entirely in
the sector $u_{4}<0$ and is endless (not ended by a fixed point).

This endless renormalized trajectory is associated with systems undergoing a
first-order phase transition. This is due to the absence of fixed point \cite
{4307}, in which case the correlation length $\xi $ cannot be made infinite
although for some systems lying close to T$_{1}^{^{\prime \prime }}$ and
attracted to it (i.e. at the transition temperature), $\xi $ may be very
large (because T$_{1}^{^{\prime \prime }}$ is endless), in which cases one
may say that the transition is almost of second order \cite{3353}. Of
course, a domain of first-order phase transition in $S$ was expected from
the usual arguments \cite{3390,4612}, we only specify better the conditions
of realization, in $S$, of the first-order transition.

Fig. \ref{fig4} shows the attractive trajectory T$_{1}^{^{\prime \prime }}$
together with the attractive tricritical line approaching the Gaussian fixed
point. The tricritical surface $S_{\text{t}}$ separates the first-order
surface $S_{\text{f}}$ from the critical surface $S_{\text{c}}$. Fig. \ref
{fig4} shows also that systems lying close to the tricritical surface may
still be attracted to the Wilson-Fisher fixed point. In this case the
effective exponents may undergo a very retarded crossover to the asymptotic
Ising values compared to usual systems corresponding to initial points
chosen in the sector $u_{4}>0$ of $S$. Fig. \ref{fig5} illustrates how minus the
inverse of (\ref{eq-OmegaEff}) provides us with different evolutions
[calculated from (\ref{eq4})] of the effective exponent $\nu _{\text{eff}%
}(\tau )$ [with $\tau \propto (T-T_{\text{c}})/T_{\text{c}}]$ according to
the initial point chosen in $S$. It is worth to explain how we have defined $%
\nu _{\text{eff}}(\tau )$.

We have seen at the end of section \ref{LPA-ERGE} that the quantity (\ref
{eq-OmegaEff}) undergoes a flat extremum (or a flat inflexion point) at an
RG eigenvalue of $f^{*}$ each time the RG flow runs along an eigendirection
in the vicinity of $f^{*}$. Now it happens that this extremum is less and
less flat as one chooses larger and larger values of $(u_{2}(0)-u_{2}^{c})$
(for the eigenvalue $\nu $) but still exists. This provides us with a way to
express the evolution of an effective exponent $\nu _{\text{eff}}$ when the
RG-substitute to $\tau $, namely $(u_{2}(0)-u_{2}^{c})/u_{2}^{c}$, is
varied. Fig. \ref{fig6} shows such an evolution for some initial Hamiltonian
(with $u_{4}(0)=4$). Notice that for such a Hamiltonian, the extremum
disappears before $\nu _{\text{eff}}$ reaches the trivial value $\frac{1}{2}$
(associated with the approach to the trivial high temperature fixed point
and to a regular ---non critical--- behavior) while in the case of a
Hamiltonian initialized close to the tricritical surface, the
classical-to-Ising crossover is complete (see fig. \ref{fig4}). This is
because in the latter case the RG trajectory comes close to the Gaussian
fixed point (and $\nu _{\text{eff}}(\tau )$ has an extremum at $\frac{1}{2}$)
before approaching to $f^{*}$. This reinforces the idea that the so-called
classical-to-Ising crossover actually exists only between the Gaussian and
Wilson-Fisher fixed points \cite{3532}.

The same configuration displayed by fig. \ref{fig4} has been obtained also
by Tetradis and Litim \cite{3552} while studying analytical solutions of an
exact RG equation in the local potential approximation for the $O(N)$-symmetric scalar theory in the large $N$
limit. But they were not able to determine ``{\sl the region in parameter
space which results in first-order transitions}'' \cite{3552}.

\subsection{Case $d=4$}

To decide whether the marginal operator (associated with the eigenvalue
equal to zero, i.e. $\lambda _{2}$ in four dimensions, or $\lambda _{3}$ in
three dimensions) is relevant or irrelevant, one must go beyond the linear
approximation. The analysis is presented in \cite{2085} for $d=4$. If one
considers a RG flow along $\chi _{2}^{+}$ such that $g_{2}(\varphi
,l)=c(l)\chi _{2}^{+}(\varphi )$, then one obtains, for small $c$: $%
c(l)=c(0)\left[ 1-Ac(0)l\right] $ with $A>0$. Hence the marginal parameter
decreases as $l$ grows. As it is well known, in four dimensions the marginal
parameter is irrelevant. However, if one considers the direction opposite to 
$\chi _{2}^{+}$ (i.e. $\chi _{2}^{-}$) then the evolution corresponds to
changing $c\rightarrow -c$. This gives, for small values of $c$: $%
c(l)=c(0)\left[ 1+A\left| c(0)\right| l\right] $ and the parameter becomes
relevant. The parameter $c$ is the renormalized $\phi ^{4}$ coupling
constant $u_{R}$ and it is known that in four dimensions the Gaussian fixed
point is IR stable for $u_{R}>0$ but IR unstable for $u_{R}<0$ \cite{399}.

We have verified that the trajectory T$_{1}^{^{\prime \prime }}$ survives
when $d=4$ (contrary to T$_{1}$, see \cite{3554}). That trajectory T$%
_{1}^{^{\prime \prime }}$ is a renormalized trajectory on which we could
define a continuum limit for the $\varphi _{4}^{4}$-field theory and if the
corresponding (perfect) action was positive for all $\varphi $, one could
say that the $\phi _{4}^{4}$-field theory with a negative coupling is
asymptotically free. Unfortunately, because the $\varphi ^{4}$-term is
dominant for large $\varphi $ in the vicinity of the origin of $S$ (due to
the relevant direction provided by $\chi _{2}^{-}$), the negative sign of
the renormalized coupling prevents the path integral to be well defined.

However, because the action to which one refers in the continuum limit (the
perfect action) is formal (because it involves an infinite number of
parameters and cannot be written down, see \cite{3554}) we wonder whether
the wrong sign of the action is actually a valid argument to reject the $%
\varphi _{4}^{4}$-field theory with a negative renormalized coupling. It is
worth to mention that the asymptotically free scalar field theory which has
recently been considered on a lattice \cite{3759} could actually be the $%
\phi _{4}^{4}$-field theory with a negative coupling to which we refer here.

{\bf Acknowledgments }We dedicate this article to Professor Yukhnovskii in
grateful recognition of his efficient and generous help in fostering the
Ukrainian-French Symposium held in Lviv in february 1993, with the hope that
in the future the contacts between our two communities will further
develop.\newpage 

\appendix 

\section{The finite difference method used}

For technical reasons, instead of studying Eq. (\ref
{eq4}), we consider the differential equation satisfied by $g(\varphi
)=f^{\prime }(\varphi )$ (i.e. the second derivative of the potential with
respect to the field):

\begin{equation}
\dot{g}=\frac{K_{d}}{2}\left[ \frac{g^{\prime \prime }}{1+g}-\frac{\left(
g^{\prime }\right) ^{2}}{\left( 1+g\right) ^{2}}\right] +\left( 1-\frac{d}{2}%
\right) \varphi g^{\prime }+2g  \label{app4}
\end{equation}

Starting with a known initial function (at ``time'' $l=0$), we follow its
evolution in $S$ by approximating the differential equation (\ref{app4}) by
finite differences and a two dimensional grid with the uniform spacings $%
dy=0.01$ and $dl=0.000390625$. The finite difference formulas for the
derivatives $g^{\prime \prime }$ and $g^{\prime }$ have been chosen with the
accuracy $O(dy^{4})$:

\begin{eqnarray}
g^{\prime }(y) &=&\frac{8}{12dy}\left[
g(y+dy)-g(y-dy)-g(y+2dy)+g(y-2dy)\right] +O(dy^{4})  \label{app'} \\
g^{\prime \prime }(y) &=&\frac{16}{12dy^{2}}\left[
g(y+dy)+g(y-dy)-30g(y)\right.  \nonumber \\
&&\left. -g(y+2dy)-g(y-2dy)\right] +O(dy^{4})  \label{app''}
\end{eqnarray}

The evolutionary function $g(y,l)$ is known (calculated) at the discret set
of points $y_{i}=i\cdot dy$ with ($i\geq 0$) and a maximum value $i_{\max
}=82$. (This value is large enough to study the approach
to the Wilson-Fisher fixed point with a great accuracy but is too small for
studying precisely the approach to the Gaussian fixed point.) At each time $%
l_{k}=k\cdot dl$, the derivatives are estimated from $%
g(y_{i},l_{k})=g(y_{i},l_{k-1})+\dot{g}(y_{i},l_{k-1})\cdot dl$ by using Eqs(%
\ref{app'},\ref{app''}) which apply only for $1<i<i_{\max -1}$. For the
marginal points $i=0,1$ we use the parity of $g(y)$ [by inserting $g(-n\cdot
dy)=g(n\cdot dy)$ for $n=1,2$ in Eqs(\ref{app'},\ref{app''})]. For the two
other marginal points $i=i_{\max -1},i_{\max }$ of the grid, there is no
fixed solution and we shall used alternately the two following conditions
[using the obvious abreviation $g(i)$ instead of $g(i\cdot dy)$]:

\begin{eqnarray*}
g^{\prime }(i) &=&g^{\prime }(i-1) \\
g^{\prime \prime }(i) &=&g^{\prime \prime }(i-1)
\end{eqnarray*}

\begin{eqnarray}
g^{\prime }(i) &=&\frac{1}{dy}\left[ \frac{25}{12}g(i)-4g(i-1)+3g(i-2)-\frac{%
4}{3}g(i-3)\right.  \nonumber \\
&&\left. +\frac{1}{4}g(i-4)+O(dy^{4})\right]  \label{app2'} \\
g^{\prime \prime }(i) &=&\frac{1}{dy^{2}}\left[ \frac{915}{244}g(i)-\frac{77%
}{6}g(i-1)+\frac{107}{6}g(i-2)-13g(i-3)\right.  \nonumber \\
&&\left. +\frac{61}{12}g(i-4)-\frac{5}{6}g(i-5)\right] +O(dy^{4})
\label{app2''}
\end{eqnarray}

Condition 2 is more accurate than condition 1 but leads sometimes to strong
unstabilities which do not appear when we first use Condition 1 and then
Condition 2 after some finite ``time'' $l_{0}$. The validity of the
procedure is tested by, for example, trying to approach a given fixed point
(see the main part of the paper).

\subsection{Appearance of spurious fixed points in the approach to the
Gausian fixed point\label{Spurious}}

In trying to determine the attractive tricritical trajectory (approaching
the Gaussian fixed point), we have encountered a spurious twice unstable
fixed point lying at some finite and non negligible distance to the Gaussian
fixed point. To understand the origin of this undesirable numerical effect,
it is necessary to discuss a bit the solution of the fixed point equation $%
\dot{f}=0$.

From (\ref{eq4}) or (\ref{app4}), one sees that the fixed point equation is
a second order non linear differential equation and a solution would be
parametrized by two arbitrary constants. One of these two constants may
easily be determined: since $g^{*}(\varphi )$ is expected to be an even function
of $\varphi $ [O(1) symmetry] then $g^{*\prime }(0)=0$ may be imposed. It
remains one free parameter, thus a one-parameter family of (nontrivial) fixed
points are solutions to the differential equation. But there is not an
infinity of physically acceptable fixed points; all but a finite number of
the solutions in the family are singular at some $\varphi _{c}$ \cite
{2085,2625,176}. Formally, by requiring the physical fixed point to be
defined for all $\varphi $ then the acceptable fixed points are limited to
the Gaussian fixed point and (for $d=3$) to the Wilson-Fisher fixed point.

However, in our study, because we numerically consider the function $g(\varphi )$ in some
finite range of values of $\varphi $ (see above: $i_{\max }=82$), it appears
that in approaching the origin of $S$, infinitely many pseudo-fixed points
exist which have there $\varphi _{c}$-singularity located outside the finite
range explicitly considered and there is at least one of them which looks
like a tricritical fixed point. When we enlarge the range of $\varphi $ to $%
i_{\max }=200$, the previously observed nontrivial tricritical fixed point
disappeared to the benefit of another one located closer to the origin. In
conclusion, a larger and larger number of grid-points must be considered as
one tries to come closer and closer to the Gaussian fixed point. This
particularity together with the slowness of the approach along a marginal
direction makes it excessively difficult to come very close to the Gaussian
fixed point.

\begin{center}
{\Large Figure captions}
\end{center}

\begin{enumerate}
\item  Evolutions for $d=3$ of the first four Hamiltonian parameters $%
u_{2}(l)$, $u_{4}(l)$, $u_{6}(l)$, $u_{8}(l)$ in a close approach to the
Wilson-Fisher fixed point $f^{*}$ along T$_1$ or T$_{1}^{^{\prime }}$. The effective inverse eigenvalue $\omega
_{\text{eff}}\left( l\right) $ is given by eq. (\ref{eq-OmegaEff}) for $%
n=2,4,6,8$. All these quantities reach the same universal value $\omega _{1}$
characteristic of the least irrelevant eigendirection of $f^{*}$. To get
this close approach to $f^{*}$ from eq. (\ref{eq4}), the initial critical
value $u_{2}^{c}$ corresponding to $u_{4}(0)=3$, has been determined with
more than twenty digits.\label{fig1}

\item  When a second condition is imposed on the initial Hamiltonian
parameters, the approach to $f^{*}$ may be adjusted such as to asymptotically
take the second least irrelevant eigendirection. Here $\omega _{\text{eff}%
}\left( l\right) $ is given by eq. (\ref{eq-OmegaEff}) for $n=2$ it clearly
undergoes (full line) a flat inflexion point at the value $\omega _{2}=2.84$
corresponding to an approach to $f^{*}$ along T$_{2}$, the greater the
critical parameter $u_{4}^{c}$ is accurately determined, the longer is the
flat extremum. Because $u_{4}^{c}$ is not completely determined [within the
available accuracy in solving eq. (\ref{eq4})] the trajectory leaves the
direction of T$_{2}$ to take one of the two directions of approach
associated to the least irrelevant inverse eigenvalue $\omega _{1}$
(corresponding to T$_{1}$ or T$_{1}^{^{\prime }}$ as indicated by dashed
curves). Here, the trajectory corresponding to the full line goes along T$%
_{1}^{^{\prime }}$. Again a flat extemum of $\omega _{\text{eff}}\left(
l\right) $ indicates the approach along an eigenvector of $f^{*}$ and
requires an accurate determination of the critical value $u_{2}^{c}$.
Because this determination is not complete, the trajectory ends up going
away from $f^{*}$ as indicated by the sudden departure of $\omega _{\text{eff%
}}\left( l\right) $ from $\omega _{1}$ for the large values of $l$.\label%
{fig2}

\item  This figure is a continuation of fig. \ref{fig2}. It shows the
various plateaux that $\omega _{\text{eff}}\left( l\right) $ undergoes along
a RG trajectory first adjusted to approach $f^{*}$ along the second
irrelevant direction (plateau at $\omega _{2}=2.84$). Because it is not
possible to determine exactly the initial conditions, the trajectory always
ends up going away the fixed point towards the trivial high temperature
fixed point characterized by the classical value $\frac{1}{2}$ (for minus
the inverse of $\omega _{\text{eff}}\left( l\right) $, thus the final
plateau at $-2$). In-between, the RG flow has been influenced by the close
vicinity of the least irrelevant eigenvector (plateau at $\omega _{1}$) and
that of the relevant eigenvector (plateau at $\omega _{0}=-\frac{1}{\nu }$).
The various regimes of the RG flows are indicated by the vertical arrows on
the left (direction of the flow with respect to the fixed point) and on the
right of the figure (distance to the fixed point).\label{fig3}

\item  Domains of attraction and repulsion of the Gaussian fixed point. The
figure represents projections onto the plane $\{u_{2},u_{4}\}$ of various RG
trajectories running in the space $S$ minus one dimension. The flows have
been obtained by solving Eq. (\ref{eq4}). Black circles represent the
Gaussian and the Wilson-Fisher (W-F FP) fixed points. The arrows indicate
the directions of the RG flows on the trajectories. The ideal trajectory
(dot line) which interpolates between these two fixed points represents the
usual renormalized trajectory T$_{1}$ corresponding to the so-called $%
\varphi _{3}^{4}$ renormalized field theory in three dimensions (usual RT
for $u_{4}>0$). White circles represent the projections onto the plane of
initial critical Hamiltonians. For $u_{4}(0)>0$, the effective Hamiltonians
run toward the Ising fixed point asymptotically along T$_{1}$ (simple
fluid). Instead, for $u_{4}(0)<0$ and according to the values of Hamiltonian
coefficients of higher order ($u_{6}$, $u_{8}$, etc.), the RG trajectories
either (A) meet an endless RT emerging from the Gaussian FP(dashed curve)
and lying entirely in the sector $u_{4}<0$ or (B) meet the usual RT T$_{1}$
to reach the Ising fixed point. The frontier which separates these two very
different cases (A and B) corresponds to initial Hamiltonians lying on the
tri-critical subspace $S_{\text{t}}$ (white square C). This is a source of
RG trajectories flowing asymptotically toward the Gaussian FPalong the
tricritical RT. Notice that the coincidence of the initial point B with the
RG trajectory starting at point A is not real (it is accidental, due to a
projection onto a plane of trajectories lying in a space of infinite dimension). 
The points A or B could correspond to the restricted primitive
model of ionic systems (see \cite{4003}).\label{fig4}

\item  Evolutions of an effective exponent $\nu _{\text{eff}}(\tau )$ [with $%
\tau \propto (T-T_{\text{c}})/T_{\text{c}}]$ along three different families
of RG trajectories (see text for additional details). The full squares
indicate the evolution of $\nu _{\text{eff}}(\tau )$ for a family of trajectories
initialized in the sector $u_{4}>0$ with $u_{4}(0)=3$ and for various values
of $u_{2}(0)$ (the same system at criticality corresponds to the white
circle ``Simple fluid'' of fig.\ \ref{fig4}). When $u_{2}(0)\rightarrow
u_{2}^{c}$ the effective exponent approaches the critical exponent value $%
\nu \cong 0.69$ compatible with the present study. One observes that the
crossover towards the classical value $\frac{1}{2}$ is not complete because $%
\nu _{\text{eff}}(\tau )$ ceases to make sense before $\tau $ becomes large.
This is not the case of the evolution represented by the full circles which
corresponds to trajectories initialized close to the Gaussian fixed point.
In this case the complete crossover reproduces the interpolation between the
Gaussian and the Wilson-Fisher fixed points and typically corresponds to the
usual answer given by field theory \cite{733}. The third evolution (full
triangles) corresponds to a family of Hamiltonian initialized close to the
tricritical surface but still attracted to the Wilson-Fisher fixed point.
One sees that the classical-to-Ising crossover is complete but highly
retarded compared to the two other cases. This is because at criticality,
the RG flow is first attracted to the Gaussian fixed point (showing then an
apparent classical value of $\nu $) before interpolating between the
Gaussian and the Wilson-Fisher fixed point.\label{fig5}

\item  Illustration of the evolution of the extrema $\nu _{\text{eff}}(l)$
[minus the inverse of eq. (\ref{eq-OmegaEff})] for various values of $\tau
=\left( u_{2}(0)-u_{2}^{c}\right) /u_{2}^{c}$ and for the family of RG flows
initialized at $u_{4}(0)=3$. The extremum (grey triangle) disappears at some
not very large value of $\tau $ (about $10^{-0.5}$) and does not reach the
classical value $\frac{1}{2}$. This induces the partial Ising-to-classical
crossover drawn on fig. \ref{fig5} (squares).\label{fig6}
\end{enumerate}

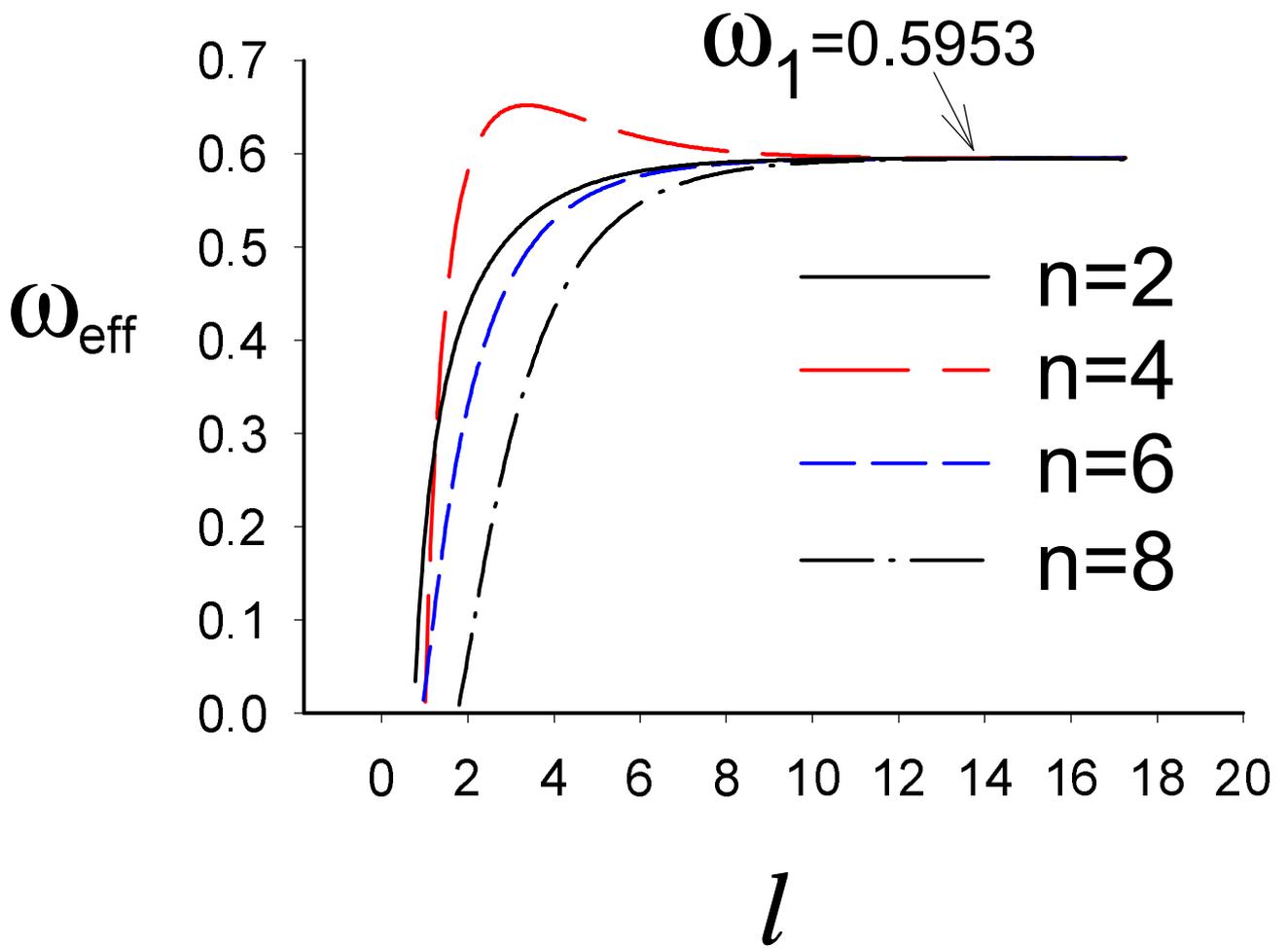
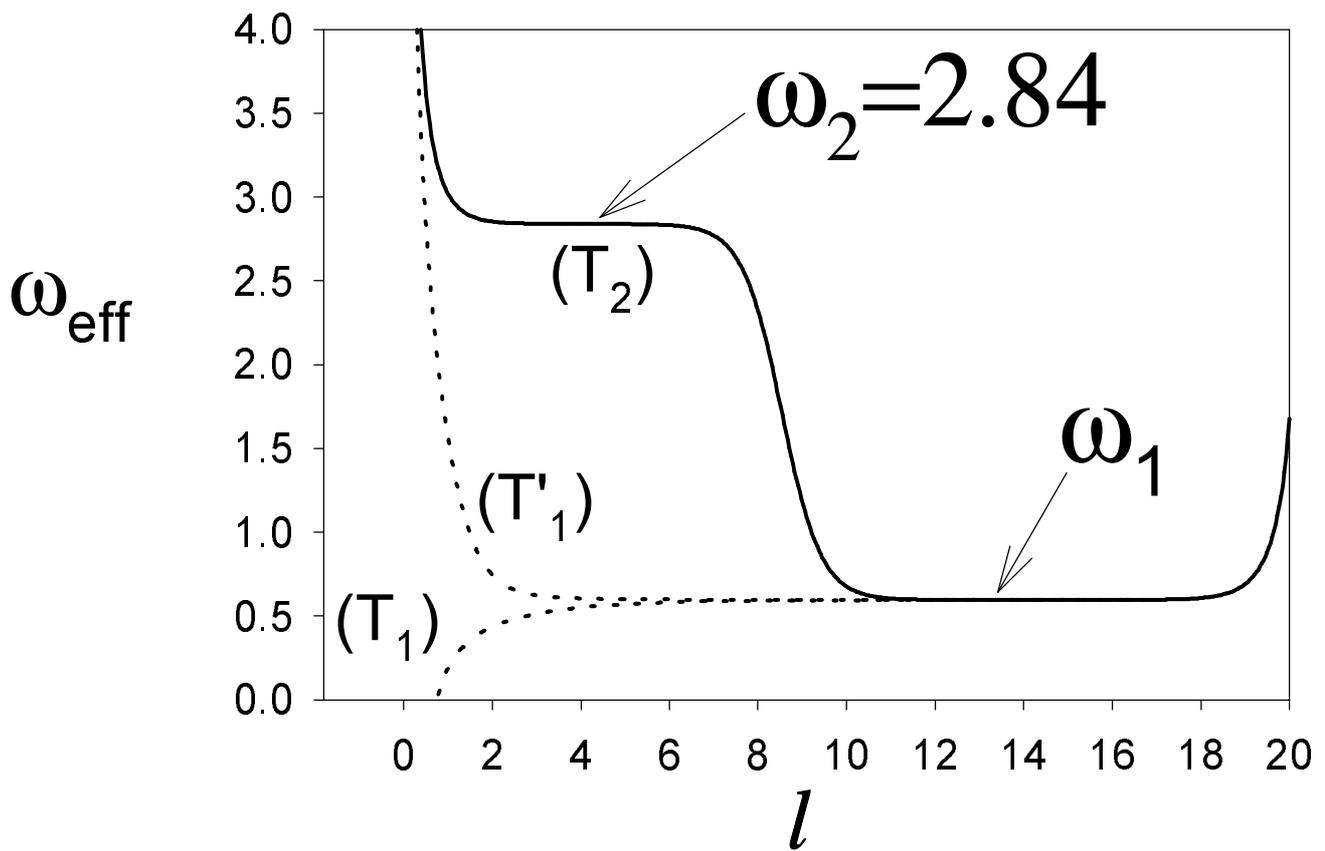
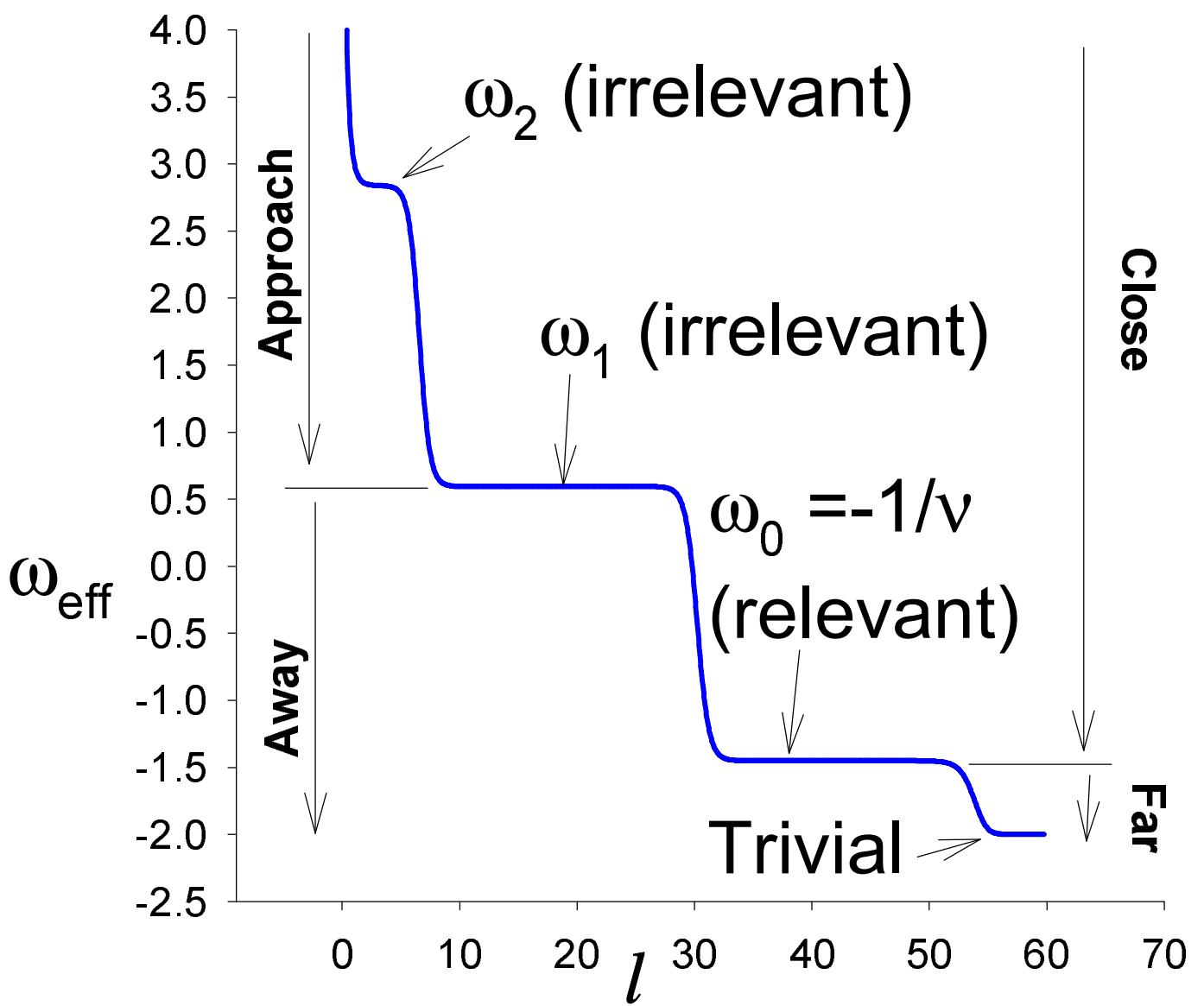
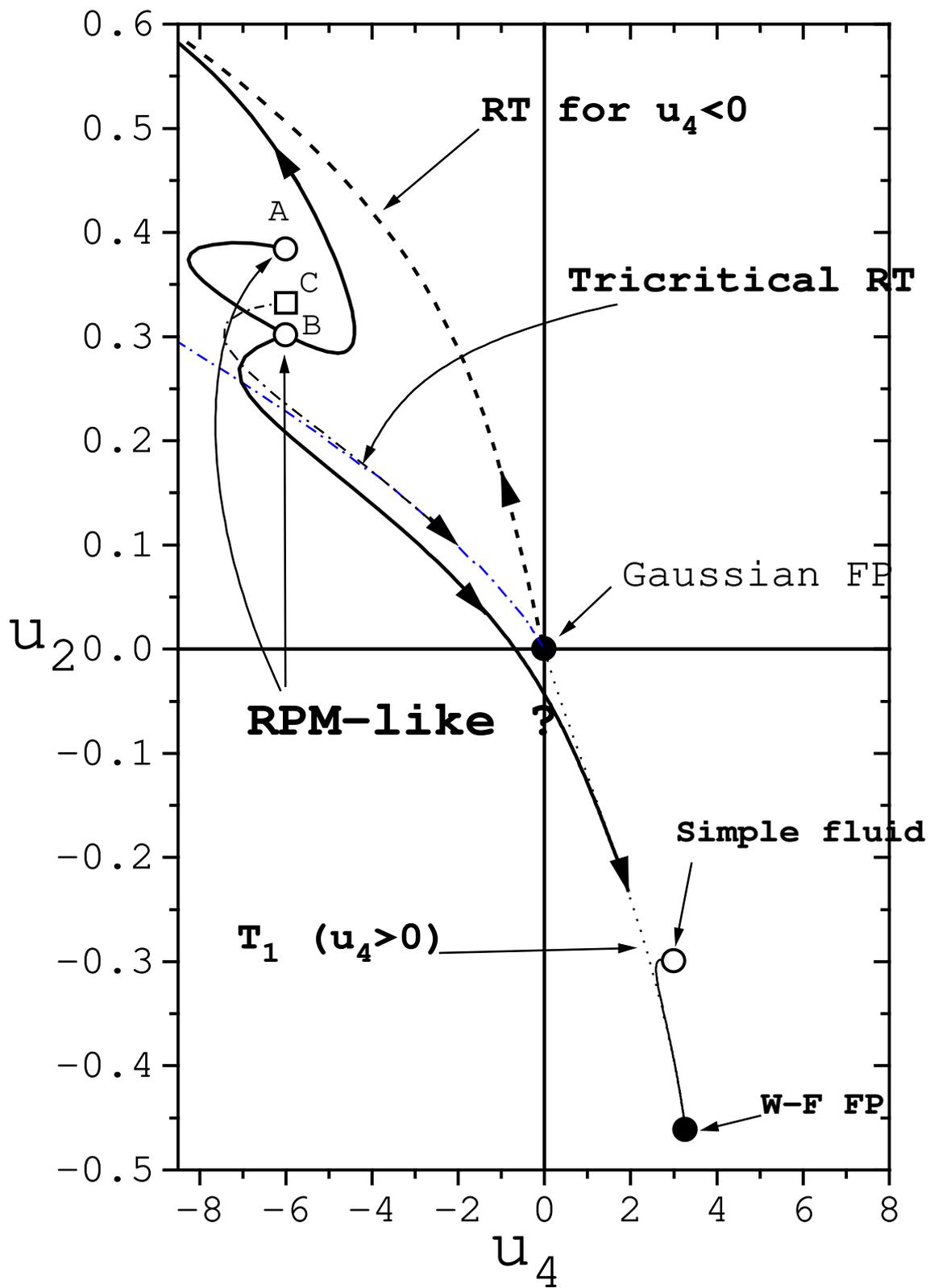
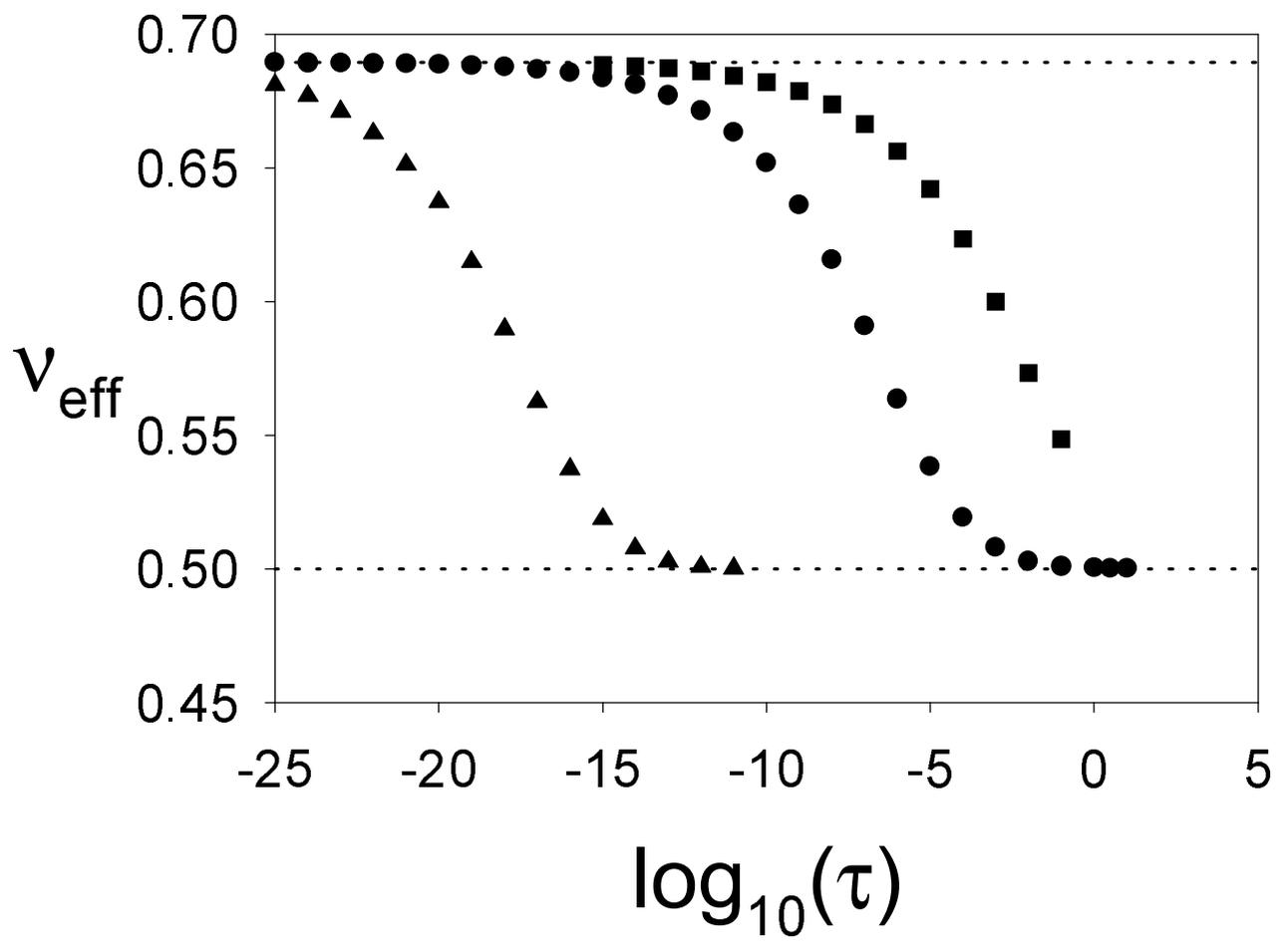
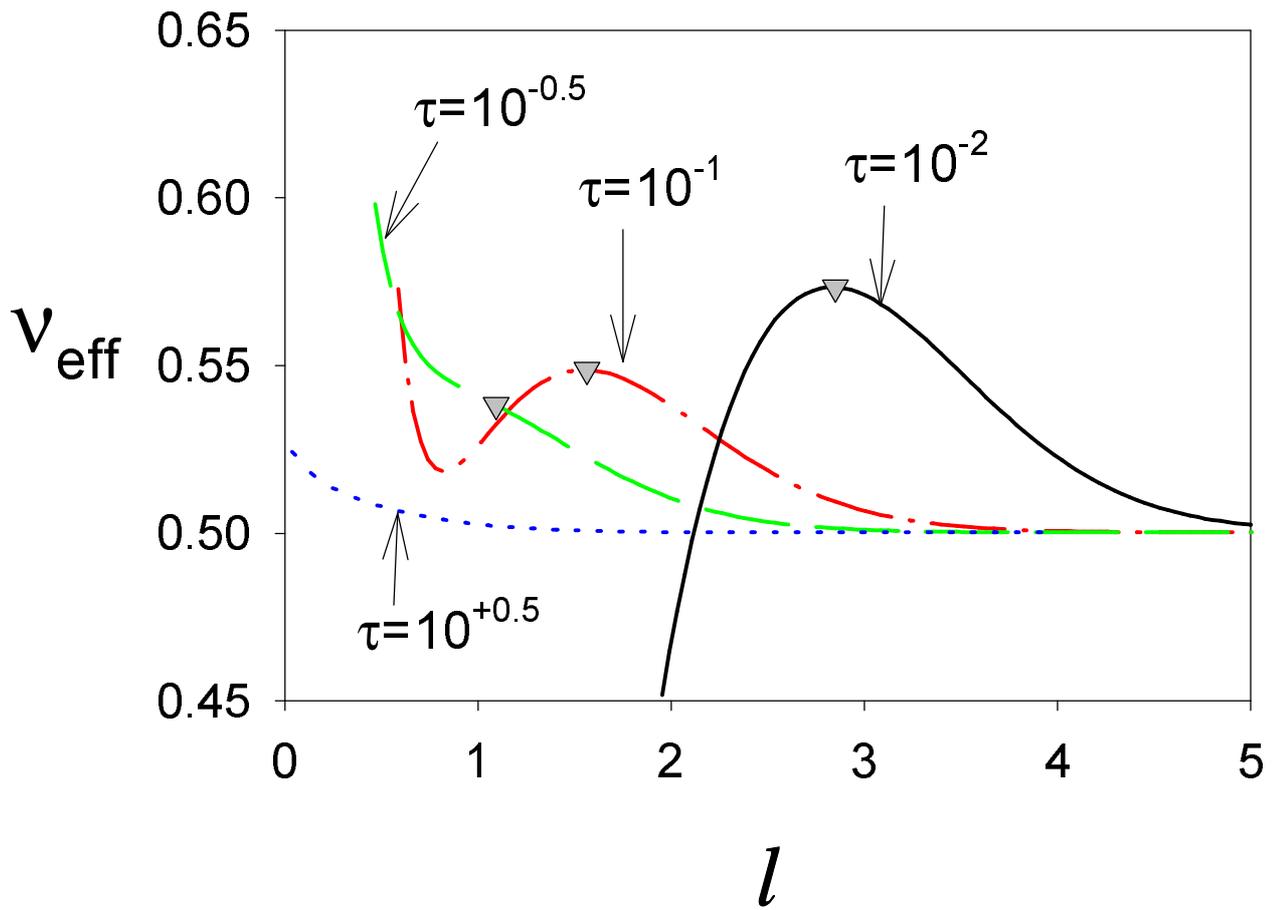
\end{document}